\documentclass[prb, twocolumn]{revtex4}

\begin{document}

\title{
Condensation of fractional excitons, non-Abelian states\\ 
in double-layer quantum Hall systems and $Z_4$ parafermions
}

\author{Edward Rezayi}
\affiliation{
Department of Physics and Astronomy,
California State University, 
Los Angeles, CA 90032
}
\author{Xiao-Gang Wen}
\affiliation{
Department of Physics, Massachusetts Institute of
Technology, Cambridge, MA 02139
}
\author{N. Read}
\affiliation{
Department of Physics,
Yale University, P.O. Box 208120, New Haven, CT 06520-8120
}

\date{{\small \today}}
\begin{abstract}
In this paper, we study a method to obtain non-Abelian FQH
state through double-layer FQH states and fractional exciton
condensation. In particular, we find that starting with the
$(330)$ double-layer state and then increasing the
interlayer tunneling strength, we may obtain a single-layer
non-Abelian FQH state $\cS(330)$.  We show that the
$\cS(330)$ state is actually the $Z_4$ parafermion
Read-Rezayi state.  We also calculate the edge excitation of
the $\cS(330)$ state.
\end{abstract}

\maketitle

\section{Introduction}

Non-Abelian fractional quantum Hall (FQH) states are a new class of FQH
states whose excitations carry non-Abelian
statistics.\cite{MR9162,Wnab,BW9215} The internal order of non-Abelian
states is so different from the symmetry breaking orders that totally
new approaches are needed to study them. Since the very
beginning, two very effective yet very different approaches were
introduced; one is based on conformal field theory\cite{MR9162} and
the other is based on projective construction.\cite{Wnab,Wpcon} Both
approaches allow us to calculate non-Abelian quasiparticles and edge
states.\cite{MR9162,BW9215,WWHopa,Wpcon,Read,Read1} The projective construction
also allows us to calculate the effective bulk topological field theory
for the non-Abelian FQH states.  This class of quantum states is
proposed to be a medium for fault tolerant quantum
computation.\cite{K032,FKL0331,DKL0252}

However, it is non trivial to realize non-Abelian states; in particular, most
observed FQH states are believed to be Abelian FQH states.  But by utilizing
the special form of electron-electron interaction in the second Landau level, a
non-Abelian FQH state, the Moore-Read Pfaffian
state,\cite{MR9162,GWW9105,Wnabhalf}
may be realized as the $\nu=5/2$ FQH state.\cite{RH0085} The $\nu=12/5$ state
may also be a non-Abelian state\cite{RR9984,RR0646}, although more
experimental studies are needed to be sure.

In this paper, we will study another possible mechanism for realizing non-Abelian
states, namely the possibility to make them in double layer
systems.  Experimentally, the hierarchal FQH states observed in double systems
are quite different from those in the first Landau level. So there
may be non-Abelian FQH states in double-layer systems.

In general, double-layer FQH states contain a class of fractional
exciton excitations.  A fractional exciton (f-exciton) is formed by a pair of
fractionally charged quasiparticle and quasihole in each layer.  Such
a fractional exciton may carry fractional statistics.  If we start
with an Abelian double-layer state, the condensation of fractional
excitons may generate a non-Abelian state.  The condensation of
fractional excitons and the resulting non-Abelian Moore-Read Pfaffian
state has been studied for the $331$ double-layer state.\cite{Wctpt}
Such a phenomenon is closely related to the transition between the
weak-pairing $p$-wave to strong pairing $p$-wave BCS
superconductors.\cite{RG0067} In this paper, we show that the
fractional exciton condensation in the $330$ double-layer state
generate the $k=4$ Read-Rezayi parafermion state.

\section{f-excitons and their condensation}

Let us consider a double-layer quantum Hall system of spin polarized electrons
at filling fraction $2/n$, where $n$ is odd.  We will consider only the $T=0$
quantum ground states.  First we assume the interlayer tunneling and
interlayer interaction to be zero. In this case the electrons in the two
layers form two independent $\nu=1/n$ Laughlin states, which is denoted as
$(nn0)$ state.  Such a state is a special case of more general $(mnl)$
double-layer states whose wave function is given by
\begin{align}
\label{mnlstate}
&\ \ \ \Phi_{mnl}(\{z_i\},\{w_i\})
\\
&\propto
\prod_{i<j} (z_i-z_j)^m(w_i-w_j)^n
\prod_{i,j} (z_i-w_j)^l e^{-\frac14 \sum_i(|z_i|^2+|w_i|^2)},
\nonumber 
\end{align}
where the complex number $z_i$ ($w_i$) are electron coordinates in the first
(second) layer.

In this paper, we will also consider the situation where $n$ is even.  In this
case, electrons must carry Bose statistics.  So the bosonic ``electrons'' will
form two independent $\nu=1/n$ Laughlin states in the two layers for even $n$.
The wave function still has the form in \eqn{mnlstate}.

Now let us still assume the interlayer tunneling to be zero, but increase the
repulsive interaction between electrons in the two different layers.  Clearly,
when the interlayer repulsive interaction is very strong, the electrons will
all stay in one of the two layers and from a single-layer $\nu=2/n$ state.
The single-layer $\nu=2/n$ state is a charge imbalanced state since the
electron densities in the two layers are different.  The charge imbalanced state
spontaneous breaks the symmetry of exchanging the two layers.  Thus the state
for small interlayer repulsion and the state for large interlayer repulsion
have different symmetries and belong to different phases.  This suggests that
as we increase interlayer repulsion from zero, we induce a quantum phase
transition at a critical interlayer repulsion. What is this quantum phase
transition?

To understand the phase transition (or the instability of the $(nn0)$ state)
induced by interlayer repulsion, let us consider the following excitation in the
$(nn0)$ state which is formed by a charge $e/n$ quasiparticle in one layer and
a charge $-e/n$ quasihole in the other layer. We will call such an excitation
a fractionalized exciton (f-exciton).  Clearly, the presence of a finite density
of f-excitons will cause a charge imbalance between the two layers.  Thus a
strong interlayer repulsion will generate many f-excitons and cause a charge
imbalanced state such as the single-layer $\nu=2/n$ state.  

This consideration
suggests that if we increase interlayer repulsion in the $(nn0)$ state, the
energy gap for the f-excitons will be reduced. When the energy gap of the
f-exciton is reduced to zero, the $(nn0)$ state will become unstable and a
quantum phase transition will happen. 
Such a quantum phase transition can be studied using the approach
developed in \Ref{Wctpt}.  The effective theory for the transition is 
found to be a Ginzburg-Landau-Chern-Simons (GLCS) theory
\begin{eqnarray}
\label{glcs}
\cL &=&  |(\prt_0 + i a_0 )\phi|^2 -  v^2 |(\prt_i + i a_i )\phi|^2
- m^2 |\phi|^2 - g |\phi|^4
\nonumber\\
&&  -\frac{n}{2} \frac{1}{4\pi} a_\mu\prt_\nu a_\la
\eps^{\mu\nu\la}.
\end{eqnarray}
Such a low energy effective theory describes
the critical point at the transition.

We would like to make a few remarks:\\
(a) Near the phase transition, both f-excitons and anti-f-excitons
become gapless. This is why the effective GLCS theory is relativistic.
The field $\phi$ describes both f-excitons and anti-f-excitons.\\
(b) The f-excitons obey fractional statistics with $\th=2\pi/n$.
The $U(1)$ gauge field $a_\mu$ with the Chern-Simons term
is introduced to reproduce the fractional statistics.\\
(c) The energy gap of the f-excitons is $m$.\\
(d) As we increase the interlayer repulsion from zero, $m^2$ will
decrease. For large interlayer repulsion, $m^2$ will be reduced to a
negative value, and cause a phase transition from the $\phi=0$ phase to
$\phi\neq 0$ phase.\\
(e) We note that in the $\phi\neq 0$ phase the densities of f-excitons and
anti-f-excitons are equal. So both the $\phi=0$ phase and $\phi\neq 0$ phases
have an equal electron density in the two layers and are charge balanced states.
The charge imbalanced state mentioned above will appear for even stronger 
interlayer repulsion.\\

\begin{figure}[tb]
\centerline{
\includegraphics[scale=0.4]{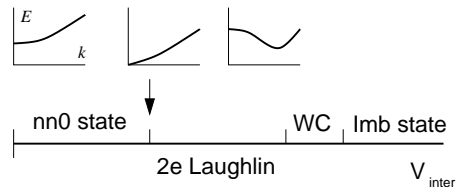}
}
\caption{
Possible dispersions of low lying charge neutral excitations.  In the
$nn0$ state, the low lying charge neutral excitations are the
f-excitons.  Such dispersion for the f-excitons leads to the
transition between the $nn0$ state and the charge $2e$ Laughlin state.
At larger $V_\text{inter}$, the Wigner crystal (WC) and a charge imbalanced state (imb state)
may appear.
}
\label{nn0NT}
\end{figure}

In the presence of inter-layer repulsion, the dispersion of the f-excitons may
have several different possible forms.  One possible dispersion relation
is illustrated in Fig. \ref{nn0NT}.  The effective theory \eq{glcs} applies to
such a situation.

\begin{figure}[tb]
\centerline{
\includegraphics[scale=0.4]{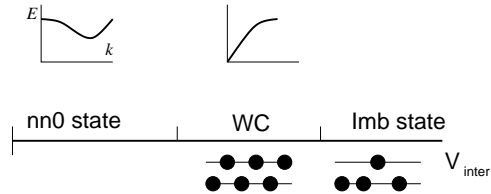}
}
\caption{
Another set of possible dispersions of low lying charge neutral excitations.
In the $nn0$ state, the low lying charge neutral excitations are
the f-excitons.
Such dispersion for the f-excitons leads to the transition between
the $nn0$ state and the WC state.
}
\label{nn0WCChImb}
\end{figure}

Another possible dispersion relation for the f-excitons
is illustrated in Fig. \ref{nn0WCChImb}.  In this case,
the effective theory \eq{glcs} does not apply.
Such case will likely lead to the WC state.

\section{Double-layer quantum Hall states
and their phase transitions
}

Let us assume that the situation described by Fig. \ref{nn0NT} is realized
and the effective theory \eq{glcs} is valid.  Clearly, the $\phi=0$ phase is
the $(nn0)$ state. What is the $\phi\neq 0$ state? In \Ref{Wctpt}, it is shown
that the $\phi\neq 0$ state is the Laughlin state for charge $2e$ electron
pairs.  The effective filling fraction for the $2e$ pairs is
$\nu_\text{eff}=1/2n$.  The $\phi=0$ to $\phi\neq 0$ transition at $m^2=0$ is
believed to be a continuous quantum phase transition.\cite{WWtran,CFW9349}
Such a transition is a transition between the $(nn0)$ double layer state and
the charge-$2e$ Laughlin state, which exists in the absence of interlayer
tunneling.

In the presence of a finite interlayer tunneling, the electron numbers in each
layer are no longer conserved separately. The effective theory \eq{glcs} will
contain extra terms to reflect this reduced symmetry.  Note that an f-exciton
is created by the operator $\phi\hat M$, where $\hat M$ is an operator that
creates $2/n$ units of $a_\mu$ flux.  Thus $\phi^n\hat M^n$ creates $n$
f-excitons, which correspond to an electron in one layer and a hole in the
other layer.  So the electron tunneling operator correspond to $\phi^n \hat
M^n + h.c.$.  Thus, with interlayer electron tunneling, the effective theory
near the transition becomes
\begin{eqnarray}
\label{glcstun}
\cL &=&  |(\prt_0 + i a_0 )\phi|^2 -  v^2 |(\prt_i + i a_i )\phi|^2
- m^2 |\phi|^2 
- g |\phi|^4
\nonumber\\
&&  
-t(\phi^n\hat M^n+h.c)
-\frac{n}{2} \frac{1}{4\pi} a_\mu\prt_\nu a_\la
\eps^{\mu\nu\la}
\end{eqnarray}
where $t$ is the amplitude of the interlayer electron tunneling.

\begin{figure}[tb]
\centerline{
\includegraphics[scale=0.4]{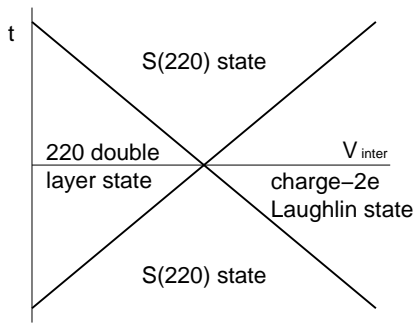}
}
\caption{
The phase diagram for the $(220)$ state, the charge-$2e$ Laughlin state,
and the $\cS(220)$ state. The $\cS(220)$ state is the Pfaffian state.
$t$ is the amplitude of the interlayer electron tunneling and $V_\text{inter}$
is the strength of interlayer repulsion.
}
\label{220}
\end{figure}

When $n=2$, the f-excitons happen to have Fermi statistics.  So the effective
theory for the $(220)$ state can be mapped to a fermion model without the
Chern-Simons term.  The fermionic effective theory is exactly soluble which
allows us to calculate the phase diagram for the $(220)$ state (see Fig.
\ref{220}).\cite{RG0067,Wctpt} For the critical point at $t=m^2=0$, the gapless
excitations are all neutral and are described by free massless Dirac fermions.
For the critical point at $t\neq 0$, the neutral gapless excitations are
described by free massless Majorana fermions.

\begin{figure}[tb]
\centerline{
\includegraphics[scale=0.4]{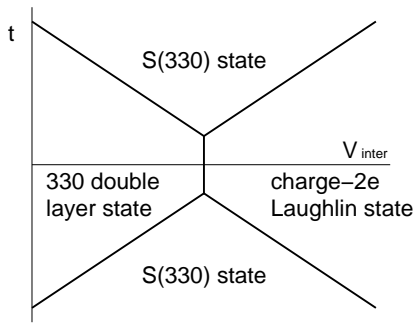}
}
\caption{
The proposed phase diagram for the $(330)$ state, 
the charge-$2e$ Laughlin state,
and $\cS(330)$ state.
}
\label{330}
\end{figure}

When $n>2$, the effective theory for the $(nn0)$ state
cannot be solved. So we do not know the phase diagram,
except that when $t=0$ we belive that there is a continuous
phase transition at $m^2=0$.  It is possible that such a
continuous phase transition is stable against a small
interlayer tunneling.  However, a large interlayer tunneling
may induce a phase transition from the $(nn0)$ state to a
new state which will be called the single-layer $(nn0)$ state
and denoted as $\cS(nn0)$ state.  This leads to the proposed
phase diagram for the $(nn0)$ state (see Fig. \ref{330}).

\section{The properties of the $\cS(nn0)$ state}

What are the properties of the $\cS(nn0)$ state? For large interlayer tunneling
$t$, the electron state created by $\psi^\dag_{e1}+\psi^\dag_{e2}$ has a lower
energy and the state created by $\psi^\dag_{e1}-\psi^\dag_{e2}$ has a higher
energy.  (Here $\psi_{e1}$ and $\psi_{e2}$ are the electron operators in the
two layers.)  So we expect that for large $t$, the $(nn0)$ state changes into
a single-layer state where electrons are always in the even state
$\psi_{e1}+\psi_{e2}$.  This consideration allows us to guess the groundstate
wave function of the $\cS(nn0)$ state (which is induced by large interlayer
tunneling from the $(nn0)$ state). Then from the guessed groundstate wave
function, we can obtain the physical properties of the $\cS(nn0)$ state.

We first note that
the wave function of the $(nn0)$ state can be expressed as
\begin{equation}
\Phi_{nn0}(\{z_i\},\{w_i\})
=\<0|\prod_i \psi_{e1}(z_i) \psi_{e2}(w_i) |nn0\>
\end{equation}
where $|0\>$ is the state with no electron and the state $|nn0\>$ is the
$(nn0)$ state.  An electron can be in a mixed even 
$\psi_{e1}+\psi_{e2}$ and odd state $\psi_{e1}-\psi_{e2}$.
After the electrons are projected into the even state
$\psi_{e1}+\psi_{e2}$, the $(nn0)$ state changes into the single-layer
$\cS(nn0)$ state.  The wave function for the $\cS(nn0)$ state is then given by
\begin{equation}
\label{Phicorr}
\Phi_{\cS(nn0)}(\{z_i\})
=\<0|\prod_i [\psi_{e1}(z_i) +\psi_{e2}(z_i)] |nn0\>
\end{equation}
For even $n$, we have
\begin{equation}
\Phi_{\cS(nn0)}(\{z_i\})
=\cS \prod_{i,j} (z_{2i}-z_{2j})^n (z_{2i+1}-z_{2j+1})^n
e^{-\frac14\sum_i |z_i|^2}
\end{equation}
and for odd $n$
\begin{equation}
\Phi_{\cS(nn0)}(\{z_i\})
=\cA \prod_{i,j} (z_{2i}-z_{2j})^n (z_{2i+1}-z_{2j+1})^n
e^{-\frac14\sum_i |z_i|^2}.
\end{equation}
Here $\cS$ is the symmetrization operator and $\cA$ the anti-symmetrization
operator. Note that $\prod_{i,j} (z_{2i}-z_{2j})^n (z_{2i+1}-z_{2j+1})^n
e^{-\frac14\sum_i |z_i|^2}$ is the wave function for the $(nn0)$ state with
$z_{2i}$ being the electron coordinates in the first layer and $z_{2i+1}$ the
coordinates in the second layer.  So the  symmetrization $\cS$ or the
anti-symmetrization $\cA$ perform the projection into even states
$\psi_{e1}+\psi_{e2}$.

Once we find an expression of the ground state wave function
in terms of correlation, such as \eqn{Phicorr},
then we can use the parton construction developed in \Ref{Wpcon}
to find the bulk effective theory and the edge excitations.
In other words, we can determine the topological order
from the expression of the ground state wave function \eq{Phicorr}.

In this paper, we will concentrate only on the edge states.  What is the
spectrum of the edge excitations of the $\cS(nn0)$ state? Let $M_0$ be the total
angular momentum of the ground state wave function $\Phi_{\cS(nn0)}$.  The
number of low-energy edge states with a fixed total number of electrons
and at angular momentum $M_0+l$ is given by $D_l$.  In appendix A, we present
a calculation of $D_l$ for the $\cS(220)$ and $\cS(330)$ state.

We find that the edge spectrum for the $\cS(220)$ state
(with an even number of electrons) is given by
\begin{align*}
D_l:\ 1, 1, 3, 5, 10, 16, 28, 43, 70 
\end{align*}
The electron operators has the following correlation
\begin{equation*}
 \<\psi_{e}(t,0) \psi^\dag_{e}(0,0)\>\sim \frac{1}{t^{g_e}} ,\ \ \ \ \ \ 
g_e=2.
\end{equation*}
We find that the minimal charged quasiparticle $\psi_q$ in
the $\cS(220)$ state
carries a charge $e/2$. 
The quasiparticle operator has the following correlation
\begin{equation*}
 \<\psi_q(t,0) \psi_q(0,0)\>\sim \frac{1}{t^{g_q}} ,\ \ \ \ \ \ 
g_q=3/8 .
\end{equation*}

The wavefunction of the $\cS(220)$ state is just the the
$\nu=1$ Pfaffian state.\cite{MR9162} The edge theory for the
$\nu=1/2$ Pfaffian state was studied numerically in \Ref{Wnabhalf}
which has an identical spectrum as the $\cS(220)$
state discussed above.

The edge spectrum for the $\cS(330)$ state (with an $4\times$
integer number of electrons) is given by
\begin{align}
\label{edgeS330}
D_l:\ 1, 1, 3, 6, 13, 23, 44, 75, 131, 215, 354, 561 ...
\end{align}
The electron operators have a correlation
\begin{equation*}
 \<\psi_{e}(t,0) \psi^\dag_{e}(0,0)\>\sim \frac{1}{t^{g_e}} ,\ \ \ \ \ \ 
g_e=3
\end{equation*}
The minimally charged quasiparticle operator $\psi_q$ carries a charge $e/6$
with a correlation
\begin{equation*}
 \<\psi_q(t,0) \psi^\dag_q(0,0)\>\sim \frac{1}{t^{g_q}} ,\ \ \ \ \ \ 
g_q=1/2 .
\end{equation*}

It turns out that the wavefunction of the $\cS(330)$ state is nothing
but the $k=4$ parafermion state introduced by Read and Rezayi.\cite{RR9984}
The edge excitations of the state are described by a charge density mode
$\rho_c$ and a $k=4$ parafermion conformal field theory.

\section{Summary}

In this paper, we discuss another route to obtain
non-Abelian FQH states through double-layer FQH states. In
particular, we propose a possibility to start with the
$(330)$ double layer state, and then increase the interlayer
tunneling strength.  We argue that such a process may change
the $(330)$ state to the $\cS(330)$ state.  Through the ideal
wave function of the $\cS(330)$ state, we find that the
$\cS(330)$ state is actually the $Z_4$ parafermion states
proposed by Read and Rezayi\cite{RR9984}(RR). We demonstrate the
equivalence of the two states in appendix B.

This research is supported by NSF grant no. DMR-0706078
(XGW), by DOE grant DE-SC0002140 (ER), and NSF grant 
DMR-0706195 (NR).

\appendix

\section{The edge excitations of the $Z_4$ Read-Rezayi parafermion state}

The edge excitations for the $(nn0)$ state can be described through
$\rho_I(x)$, $I=1,2$,which are the 1D electron densities on the edge.  $\rho_1$ is the
electron density for the first layer and $\rho_2$ for the second layer.
The Hilbert space and the dynamics of the edge excitations
are described by the following
current algebra (in the $k$-space)
\begin{align}
\label{calg}
 [\rho_{Ik},\rho_{Jk'}]=\frac{1}{n} \frac{1}{2\pi} k\del_{k+k'}\del_{IJ}
\end{align}
and the Hamiltonian
\begin{equation}
\label{Hnn0}
 H=\sum_{k>0} V_{IJ}\rho_{I,-k}\rho_{J,k}.
\end{equation}
\eqn{calg} and \eqn{Hnn0} provide a complete description of the edge
excitations. Note that \eqn{calg} and \eqn{Hnn0} just describe a collection of
harmonic oscillators labeled by $k>0$ and $I=1,2$, with the lowering operator
$a_{Ik}\propto \rho_{Ik}$ and raising operator $a^\dag_{Ik}\propto
\rho_{I,-k}$.

The electron operators on the edge are given by
\begin{equation*}
 \psi_{eI}=e^{\imth n \phi_I(x)}
\end{equation*}
where $\frac{1}{2\pi}\prt_x \phi(x)=\rho_{I}(x)$.
$\phi_{e1}$ is for the first layer and $\phi_{e2}$ the second layer.
The electron operators have the following correlation
\begin{equation*}
 \<\psi_{eI}(t,0) \psi^\dag_{eI}(0,0)\>\sim \frac{1}{t^{g_e}} ,\ \ \ \ \ \ 
g_e=n.
\end{equation*}

After we obtain the edge theory for the $(nn0)$ state and
identify the electron operator, we are ready to do the
projection into the even states and to obtain the edge
theory for the $\cS(nn0)$ state.  To do so, we first identify
a new electron operator $\psi_e$ and then use the new
electron operator and only the new electron operator to
create the edge excitations of the single-layer $\cS(nn0)$
state.

Since the $\cS(nn0)$ state only contains electrons in the even state, so
the new electron operator is
\begin{equation*}
 \psi_e(x)=\psi_{e1}(x)+\psi_{e2}(x).
\end{equation*}
The other combination $\psi_{e1}(x)-\psi_{e2}(x)$ does not
generate gapless edge excitations and is dropped.  If we use
$\psi_{e1}$ and $\psi_{e2}$ to generate gapless edge
excitations, we will generate all the edge excitations of
the $(nn0)$ state.  However, to obtain the gapless edge
excitations for the $\cS(nn0)$ state, we can only use
$\psi_e=\psi_{e1}+\psi_{e2}$.  So the $\cS(nn0)$ state will have
fewer gapless edge excitations.

What kind of edge excitations does $\psi_e$ generate?
To answer such a question, we introduce
the total electron density $\rho_c$ and
the relative electron density $\rho_r$ of the two layers:
\begin{equation*}
 \rho_c=\rho_1+\rho_2,\ \ \ \  \ \ \
 \rho_r=\rho_1-\rho_2 .
\end{equation*}
$\rho_c$ and $\rho_r$ satisfy the following current algebra
\begin{align*}
 [\rho_{ck},\rho_{ck'}]&=\frac{2}{n} \frac{1}{2\pi} k\del_{k+k'}
\nonumber\\
 [\rho_{rk},\rho_{rk'}]&=\frac{2}{n} \frac{1}{2\pi} k\del_{k+k'}
\end{align*}
In terms of $\phi_c$ and $\phi_r$, defined through
$\frac{1}{2\pi}\prt_x\phi_c=\rho_c$ and
$\frac{1}{2\pi}\prt_x\phi_r=\rho_r$, the new electron operator has the form
\begin{equation*}
\psi_e=e^{\imth n\phi_c/2} \cos(n\phi_r/2).
\end{equation*}
From the relation between the FQH wave function and
CFT,\cite{MR9162,WWHopa} the correlation of the above
electron operator reproduces the $\cS(nn0)$ wave function
\begin{equation}
\Phi_{\cS(nn0)}(\{z_i\})=\<
V(z_\infty) \prod_i \psi_e(z_i)
\> .
\end{equation}

First, let us consider the case for $n=3$.  We have shown (see appendix B)
that $\Phi_{\cS(330)}(\{z_i\})$ is the $Z_4$ Read-Rezayi
parafermion state.  This means that $\Phi_{\cS(330)}(\{z_i\})$
can be expressed as a correlation function in the $Z_4$
parafermion CFT:
\begin{equation}
\Phi_{\cS(330)}(\{z_i\})=\<
V(z_\infty) \prod_i \psi_1(z_i)e^{\imth 3\phi_c(z_i)/2}
\> ,
\end{equation}
where $\psi_1$ is the simple current operator that generates
the $Z_4$ parafermion CFT.  We see that 
\begin{equation}
 \psi_e(z)=\psi_1(z)e^{\imth 3\phi_c(z)/2}
\end{equation}
and we can identify $\cos(3\phi_r/2)$ with $\psi_1$.  In
fact, $\cos(3\phi_r/2)$ has a scaling dimension $3/4$ which
matches that of $\psi_1$.

The operator product expansion of the
$\psi_e(z)=\psi_1(z)e^{\imth n\phi_c(z)/2}$ will generate
the operator $\rho_c$, $\psi_1\psi_1^\dag\sim \psi_1
(\psi_1)^3$ etc.  So the edge excitations (with a fixed
total electron number) form a Hilbert space $\cH$ which is
the direct product of two Hilbert spaces:
$\cH=\cH_{U(1)}\otimes \cH_{Z_4}$.  $\cH_{U(1)}$ is
generated by $\rho_c$ and $\cH_{Z_4}$ is generated by $
\psi_1(z_1) \psi_1(z_2) \psi_1(z_3) \psi_1(z_4) $.

What is the spectrum of the edge excitations generated by
$\rho_c$ and $\psi_1(z_1) \psi_1(z_2) \psi_1(z_3)
\psi_1(z_4) $?
Let $M_0$ be the total angular momentum of the ground state
wave function $\Phi_{\cS(330)}$.
The number of the low-energy edge states at angular momentum
$M_0+l$ is given by $D_l$. We can introduce a function
\begin{equation*}
 \text{ch}(\xi)=\sum_{l=0}^\infty D_l\xi^l
\end{equation*}
to describe the edge spectrum $D_l$. The function
is called the character of the edge excitations.

If we only use $\rho_c$ to generate edge excitations, the character will be
\begin{equation*}
 \text{ch}_c(\xi)={\frac{1}{\prod_{l=1}^\infty (1-\xi^l)}}.
\end{equation*}
If we only use $\psi_1(z_1) \psi_1(z_2) \psi_1(z_3)
\psi_1(z_4) $ to generate edge excitations, the character
will be that of $Z_4$ parafermion CFT.\cite{DQ} The character
(in the vacuum sector) for the $Z_k$ parafermion CFT is
given by
\begin{align}
& \text{ch}_{Z_k}(\xi) =[\text{ch}_c(\xi)]^2\times
\\
& \sum_{r,s=0}^{\infty} (-1)^{r+s} \xi^{r(r+1)+s(s+1)+rs(k+1)}
\Big[
1-\xi^{(k+1)(1+r+s)}
\Big].
\nonumber 
\end{align}
If  we apply both $\rho_c$ and $\psi_1(z_1) \psi_1(z_2)
\psi_1(z_3) \psi_1(z_4) $ to generate edge excitations, the
character will be
\begin{align}
 \text{ch}(\xi)&= \text{ch}_c(\xi) \text{ch}_{Z_4}(\xi).
\end{align}
The character $\text{ch}(\xi)$ describes the edge spectrum
of the $\cS(330)$ or $Z_4$ Read-Rezayi parafermion state.  

We find that the edge spectrum for the $\cS(330)$ state 
is given by
\begin{align*}
D_l:\ 1, 1, 3, 6, 13, 23, 44, 75, 131, 215, 354, 561 ...
\end{align*}
The electron operators have the following correlation
\begin{equation*}
 \<\psi_{e}(t,0) \psi^\dag_{e}(0,0)\>\sim \frac{1}{t^{g_e}} ,\ \ \ \ \ \ 
g_e=n.
\end{equation*}
Thus $g_e=3$ for the $\cS(330)$ state.

A quasiparticle operator $\psi_q$ must satisfy the following condition:
in the operator product between $\psi_q$ and $\psi_e$
\begin{equation*}
 \psi_e(x)\psi_q(0)=\sum_i \frac{1}{x^{\al_i}} O_i(0),
\end{equation*}
where the exponents $\al_i$ must all be integers.  In this case, we say that the
quasiparticle operator $\psi_q$ is mutually local with respect to the electron
operator $\psi_e$.\cite{Wtop} 
One of the quasiparticles in the $\cS(nn0)$ state
is created by 
\begin{equation*}
\psi_q^{(\frac12,\frac12)}=e^{\imth \phi_c/2} \cos(\phi_r/2)
\end{equation*}
which carries charge $e/n$. 
The quasiparticle operator has the following correlation
\begin{equation*}
 \<\psi_q^{(\frac12,\frac12)}(t,0) \psi^{(\frac12,\frac12)\dag}_q(0,0)\>\sim \frac{1}{t^{g_q}} ,\ \ \ \ \ \ 
g^{(\frac12,\frac12)}_q=1/n .
\end{equation*}
For the $Z_4$ state, we find $g^{(\frac12,\frac12)}_q=1/3$
and $Q^{(\frac12,\frac12)}_q=e/3$.

Similarly $\psi_q^{(k,l)}$ defined by
\begin{equation*}
\psi_q^{(k,l)}=e^{\imth k \phi_c} \cos[l\phi_r],\ \ \ \
2k=\text{integer},\ \ 
2l=\text{integer},
\end{equation*}
are also valid quasiparticle operators, which carry charge
$Q^{(k,l)}=2ke/n$. The exponent for such a quasiparticle
is given by $g^{(k,l)}_q=\frac{1}{2n}(k^2+l^2)$.

However, the quasiparticle $\psi_q^{(\frac12,\frac12)}$ is
not the one that carries minimal charge. The minimally
charged quasiparticle has a charge $Q_q=e/6$ and a scaling
dimension $h_q=\frac{1}{16}+\frac{1}{48}=\frac{1}{4}$.  So
the exponent for such a quasiparticle is $g_q=2h_q=1/2$.

Now, let us consider the case for $n=2$.  We can fermionize
the $\rho_r$ sector. Introducing a complex fermion field
$\psi(x)$, the states generated by $\rho_r(x)$ can be equally
generated by $\psi^\dag(x)\psi(x)$.  In the fermion
description $e^{i 2\phi_r(x)}\sim \psi(x)$.  Thus
$\cos[2\phi_r(x)] \sim \psi(x)+\psi^\dag(x)\equiv \la(x)$
where $\la(x)$ is a Majorana fermion field which satisfies
\begin{equation*}
 \la(x)\la(0)=\frac{1}{x} +x [\prt \la(0)]\la(0)+\cdots
\end{equation*}
Using $\psi_e=e^{\imth 2 \phi_c}\la$, we find that
\begin{align*}
\psi_e(x)\psi_e^\dag(0)
&= 
x^{-2}+\imth 4\pi \rho_c (0) x^{-1} 
\nonumber\\&\ \ \ 
- 8\pi^2 \rho_c^2(0) -\la(0)\prt \la(0) +\cdots
\end{align*}
Thus the edge excitations for the $n=2$ case are generated by
$\rho_c$ and $\la \prt \la$.  These edge excitations are
described by a density mode $\rho_c$ and a Majorana fermion
$\la$.\cite{Wnabhalf}

If we only use $\rho_c$ to generate edge excitations, the
character will be
\begin{equation*}
 \text{ch}_c(\xi)={\frac{1}{\prod_{l=1}^\infty (1-\xi^l)}}.
\end{equation*}
If we only use $\la \prt \la $ to generate edge excitations,
the character will be $\text{ch}_{Z_2}(\xi).$ If  we apply
both $\rho_c$ and $\la \prt \la  $ to generate edge
excitations, the character will be
\begin{align}
 \text{ch}(\xi)&= \text{ch}_c(\xi) \text{ch}_{Z_2}(\xi).
\end{align}
We find that\cite{Wnabhalf}
\begin{align*}
D_l:\ 1, 1, 3, 5, 10, 16, 28, 43, 70, 105, 161, 236 ...
\end{align*}

\section{Relation to the $Z_4$ RR state}

To show the equivalence the of ${\cS}(330)$ and $Z_4$ RR state we will compare the boson 
version of the two states by dividing out a Jastrow factor $\prod_{i<j)}(z_i-z_j)$ from the
former.
 We will invoke the well known property of the $z_k$ RR states: as a function of their 
complex coordinates the 
wave functions vanish quadratically 
when $k+1$ 
particles approach. We will also assume that ${\cS}(330)$ does not vanish identically which,
as shown below, is not the case for $N/2$ odd.  This is to be expected since the RR state for
$k=4$ does not exist for odd $N/2$.  
Starting with ${\cS}(330)$ we write it as
\begin{align}
\label{eq:A330}
&\ \ \
\Psi_{{\cS}(330)}(z_1,z_2,\ldots,z_N)
\\
&=\sum_{Q}(-)^Q \prod_{Qi<Qj} (z_{Qi}-z_{Qj})^3
\prod_{Qm<Qn}(z_{Qm}-z_{Qn})^3
\nonumber 
\end{align}
where $Q$ is a permutation of $N$ objects, and $i\ \mbox{and}\ j\ne\ m\ \mbox{or}\ n$,
We have omitted the exponential factors. We have used the antisymmetry of the Laughlin 
terms  to restrict the number of permutations to a minimum.  We will let 5 particles 
approach each other while keeping the other $N-5$ well separated from this group and each 
other.  We relabel the coordinates $\omega_1$, $\omega_2, \ldots, \omega_5$. Then in 
\eqn{eq:A330}
there are 5 types of terms depending on how these 5 particles are distributed in the two 
Laughlin factors: $[5,0]$ (and $[0,5]$), $[4,1]$, and $[3,2]$.  It can be seen that the
total number of such terms respectively are: 
\[2{N-5\choose{N/2}},\  2{N-5\choose{N/2-1}}{5\choose{4}},\ \mbox{and}\ 2{N-5\choose{N/2-2}}
{5\choose{3}}.\]
It is straightforward to show that the sum of the above is ${N\choose{N/2}}$, 
which is the number of terms in \eqn{eq:A330}.

First consider the $[5,0]$ terms.  After dividing out the Jastrow factor it is clear that
these terms vanish if any two of the five particles coordinates are equal. For the $[4,1]$ 
terms,  a little algebra shows that they vanish if 4 of the coordinates are set to be equal.
We are left with $[3,2]$ terms which contain factors such as
\begin{equation}
(\omega_1-\omega_2)^3 (\omega_1-\omega_3)^3 (\omega_2-\omega_3)^3 (\omega_4-\omega_5)^3.
\end{equation}
In the rest of the wavefunction we set the $\omega_i$'s to their common values $\omega$
without loss of generality.  We can then collect the terms in $Q$ that permute the five
particles amongst themselves (${5\choose{3}}=10$ terms). Dividing out the Jastrow factor, we
obtain
\begin{equation}
-3\sum_{i<j}^5 (\omega_i-\omega_j)^2.
\end{equation}
We have included an equal expression 
from the $[2,3]$ 
distribution of the $\omega's$. To see that $[3,2]$ and $[2,3]$ are  identical 
for even $N/2$ consider the two permutations that interchange the Laughlin 
terms:
\begin{eqnarray*}
Q=I: (1,2,3,\ldots, N/2;N/2+1,N/2,+2,\ldots,N)\mbox{,} \\
Q^\prime:(N/2+1,N/2+2,\ldots,N:1,2,\ldots,N/2).
\end{eqnarray*}
Clearly, these permutations produce identical terms but,
depending on whether $N/2$ is even/odd, they have the
same/opposite parities.  Since every permutation can be
paired in this way, it can be seen that exchanging the
Laughlin factors preserves the overall sign for even $N/2$,
leading to the equality of $[3,2]$ and $[2,3]$.  On the
other hand, for odd $N/2$ with opposite signs,
\eqn{eq:A330} identically vanishes.  

So far we have established that the two functions are
equivalent but this still leaves out an overall
multiplicative factor.  We have empirically determined that
the factor is 
\begin{align}
{ \Psi_{RR}(z_1,z_2,\ldots,z_N)\over \Psi_{{\cS}(330)}(z_1,z_2,\ldots,z_N)}
=
(-1)^{N/4(N/4+1)\over 2} 4^{N/4},
\end{align}
where we have used the following form for the $Z_4$ wavefunction:
\begin{align}
&\ \ \ \Psi_{RR}(z_1,z_2,\ldots,z_N)
\nonumber\\
&=\sum_{Q} \prod_{Qi<Qj} (z_{Qi}-z_{Qj})^2
\prod_{Qk<Ql}(z_{Qk}-z_{Ql})^2 \times
\nonumber\\
& \ \ \ \ \ \ \ \
\prod_{Qm<Qn}(z_{Qm}-z_{Qn})^2 \prod_{Qr<Qp}(z_{Qr}-z_{Qp})^2
\end{align}


\end{document}